\documentclass[preprint ]{aastex}

\received{}
 \revised{}
\accepted{}
\journalid{}{}
\articleid{}{}

\shorttitle{Quasi-Periodic Oscillations}
\shortauthors{Mauche}

\slugcomment{Accepted for publication in {\it Ap.J.\/} 2002 July 23}

\newcommand{\Mdot}{\dot{M}}
\newcommand{\Msun}{{\rm M_{\odot}}}
\newcommand{\lax}{{\lower0.75ex\hbox{ $<$ }\atop\raise0.5ex\hbox{ $\sim$ }}}
\newcommand{\gax}{{\lower0.75ex\hbox{ $>$ }\atop\raise0.5ex\hbox{ $\sim$ }}}

\begin{document}

\title{Correlation of the Quasi-Periodic Oscillation Frequencies \\
       of White Dwarf, Neutron Star, and Black Hole Binaries}

\author{Christopher W.\ Mauche}
\affil{Lawrence Livermore National Laboratory,
       L-43, 7000 East Avenue, Livermore, CA 94550; \\
       mauche@cygnus.llnl.gov}



\begin{abstract}

Using data obtained in 1994 June/July with the {\it Extreme Ultraviolet
Explorer\/} deep survey photometer and in 2001 January with the {\it
Chandra X-ray Observatory\/} Low Energy Transmission Grating Spectrograph,
we investigate the extreme-ultraviolet (EUV) and soft X-ray oscillations
of the dwarf nova SS~Cyg in outburst. We find quasi-periodic oscillations
(QPOs) at $\nu_0\approx 0.012$ Hz and $\nu_1\approx 0.13$ Hz in the EUV
flux and at $\nu_0\approx 0.0090$ Hz, $\nu_1\approx 0.11$ Hz, and possibly
$\nu_2\approx\nu_0 + \nu_1\approx 0.12$ Hz in the soft X-ray flux. These
data, combined with the optical data of Woudt \& Warner for VW~Hyi,
extend the Psaltis, Belloni, \& van der Klis $\nu_{\rm high}$--$\nu_{\rm
low}$ correlation for neutron star and black hole low-mass X-ray binaries
(LMXBs) nearly two orders of magnitude in frequency, with $\nu_{\rm low}
\approx 0.08\, \nu_{\rm high}$. This correlation identifies the
high-frequency quasi-coherent oscillations (so-called ``dwarf nova
oscillations'') of cataclysmic variables (CVs) with the kilohertz QPOs
of LMXBs, and the low-frequency QPOs of CVs with the horizontal branch
oscillations (or the broad noise component identified as such) of LMXBs.
Assuming that the same mechanisms produce the QPOs in white dwarf,
neutron star, and black hole binaries, we find that the data exclude
the relativistic precession model and the magnetospheric and sonic-point
beat-frequency models (as well as {\it any\/} model requiring the
presence or absence of a stellar surface or magnetic field); more
promising are models that interpret QPOs as manifestations of disk
accretion onto any low-magnetic field compact object.

\end{abstract}

\keywords{accretion, accretion disks ---
          novae, cataclysmic variables ---
          stars: individual (SS~Cygni) ---
          stars: neutron ---
          stars: oscillations ---
          X-rays: binaries}

 

\section{Introduction}

Rapid periodic oscillations are observed in the optical flux of high
accretion rate (``high-$\Mdot $'') cataclysmic variables (CVs) (nova-like
variables and dwarf novae in outburst) \citep{pat81, war95a, war95b}.
These oscillations have high coherence ($Q\approx 10^4$--$10^6$), short
periods ($P\approx 7$--40 s), low amplitudes ($A\lax 0.5$\%), and are
sinusoidal to within the limits of measurement. They are referred to as
``dwarf nova oscillations'' (DNOs) to distinguish them from the longer
period, low coherence ($Q\approx 1$--10) quasi-periodic oscillations
(QPOs) of high-$\Mdot $ CVs, and the longer period, high coherence ($Q
\approx 10^{10}$--$10^{12}$) oscillations of DQ~Her stars. DNOs appear on
the rising branch of the light curve of dwarf nova outbursts, typically
persist through maximum, and disappear on the declining branch of the
light curve. The period of the oscillation decreases on the rising branch
and increases on the declining branch, but because the period reaches
minimum about one day after maximum optical flux, dwarf novae describe
loops in plots of oscillation period versus optical flux.

Rapid periodic oscillations have been detected in the flux of numerous 
high-$\Mdot $ CVs, but the dwarf nova SS~Cyg in outburst is one of the
best studied. Oscillations in the optical flux have been detected with
periods ranging from 7.3 s to 11 s \citep{pat78, hor80, hil81, pat81},
and oscillations in the extreme-ultraviolet (EUV) and soft X-ray flux
have been detected with periods ranging from 2.8 s to 11 s in data from
{\it HEAO 1\/}, {\it EXOSAT\/}, the {\it Extreme Ultraviolet Explorer\/}
({\it EUVE\/}), and {\it ROSAT\/} \citep{cor80, cor84, jon92, mau96,
tes97,mau01b, mau02a}. \citet{mau96, mau02a} showed that the EUV
oscillation period is a single-valued function of the EUV flux, explained
the loops observed in plots of oscillation period versus optical flux as
the result of the delay between the rise of the optical and EUV flux at
the beginning of dwarf nova outbursts \citep{mau01a}, and so provided
strong evidence for the proposal, first mooted by \citet{pat81}, that the
oscillation period depends solely on the mass-accretion rate onto the
white dwarf. \citet{mau01b} observed ``frequency doubling'' of the EUV
oscillation of SS~Cyg during the rising branch of its 1996 October
outburst, demonstrated that the optical and EUV oscillation periods are
equal, and that their relative phase delay is consistent with zero.

In this communication, we present an analysis of observations of SS~Cyg
in outburst obtained in 1994 June/July with the {\it EUVE\/} deep
survey (DS) photometer and in 2001 January with the {\it Chandra X-ray
Observatory\/} Low Energy Transmission Grating Spectrograph (LETGS). In
\S 2 we present the observations and data analysis, finding that the
EUV flux was oscillating at frequencies $\nu_0\approx 0.012$ Hz and
$\nu_1 \approx 0.13$ Hz, while the soft X-ray flux was oscillating
at frequencies $\nu_0\approx 0.0090$ Hz, $\nu_1\approx 0.11$ Hz, and
possibly $\nu_2\approx\nu_0+\nu_1\approx 0.12$ Hz. In \S 3 we discuss
these results in the context of similar results from optical observations
of the dwarf nova VW~Hyi in outburst and X-ray observations of neutron
star and black hole low-mass X-ray binaries (LMXBs). As first pointed
out by \citet{war02a, war02b}, white dwarf, neutron star, and black hole
binaries share a common correlation in the frequencies of their QPOs,
extending over nearly five orders of magnitude in frequency. The
implications of this result for QPO models are discussed, we discuss how
additional observations of CVs can provide unique and quantitative tests
of QPO models, and close in \S 4 with a summary.

\section{Observations and Data Analysis}

\subsection{Chandra LETGS}

Our pre-approved target-of-opportunity {\it Chandra\/} Low Energy
Transmission Grating (LETG) and High Resolution Camera (HRC) observation
of SS~Cyg was performed between 2001 January 16 $\rm 21^h13^m$ and
January 17 $\rm 10^h50^m$ UT during the plateau phase of a wide normal
(asymmetric) dwarf nova outburst that began on January 12. When the
observation began, SS~Cyg had been at maximum optical light ($V\approx
8.5$) for approximately 3 days; long enough, given the delay of 1.5
days between the rise of the optical and EUV light curves of SS~Cyg
\citep{mau01a}, for the EUV/soft X-ray flux to have reached maximum and
for the system to have reached a quasi-steady-state. Indeed, the zero-
and $\pm$ first-order LETG/HRC count rates were observed to be roughly
constant throughout the 47.3 ks observation at 1.5 and 4.9 $\rm
counts~s^{-1}$, respectively.

The files used for this analysis were created on 2002 January 7 by the
pipeline data reduction software using CIAO 2.0 with ASDCS version 6.5.1.
Events were extracted from the level 2 evt file, and the mean spectrum
and source and background region masks were extracted from the level 2
pha file. The mean LETG spectrum contains two distinct components: (1)
a hard component shortward of 40 \AA \ consisting of a bremsstrahlung
continuum and emission lines of H- and He-like C, N, O, Ne, Mg, and Si
and L-shell Fe (most prominently \ion{Fe}{17}) and (2) a soft component
extending from 42 \AA \ to 120 \AA\ that appears to consist of a continuum
significantly modified by a forest of emission and absorption features
(possibly even P~Cygni profiles). In the 72--120 \AA \ wave band, the LETG
spectrum is essentially identical to that measured by the {\it EUVE\/}
short wavelength (SW) spectrometer during the 1993 August \citep{mau95}
and 1996 October \citep{whe02} outbursts of SS~Cyg. These hard and soft
components of the X-ray spectrum of SS~Cyg are understood to be due to
the optically thin and optically thick portions of the boundary layer
between the disk and the surface of the white dwarf.

To investigate the temporal properties of the hard and soft X-ray flux
of SS~Cyg, we processed the $\pm $ first-order LETG events as follows.
First, we created a list of source and background events by filtering the
events with the source and background region masks. Second, we applied a
wavelength filter that excluded events near the HRC chip gaps: for the
positive order, events with wavelengths $\lambda =60.0$--67.5 \AA \ were
excluded, while for the negative order, events with wavelengths $\lambda
=49.8$--57.2 \AA \ were excluded. Without this filter, a portion of the
soft component of the spectrum moves on and off the HRC chips, and so
produces signals in the power spectra at the spacecraft dither periods of
707 s and 1000 s. Third, various wavelength cuts were applied to isolate
the hard ($\lambda =1$--42 \AA ) component of the spectrum, the soft
($\lambda = 42$--120 \AA ) component of the spectrum, and various subsets
thereof (see below). Events passing these filters were used to create
background-subtracted light curves with 1 s time resolution, and power
spectra of these light curves were examined for evidence of periodic flux
modulations. Consistent with more sensitive searches by {\it HEAO-1\/}
\citep{swa79}, {\it ASCA\/} \citep{nou94}, {\it Ginga\/} \citep{pon95},
and the {\it Rossi X-ray Timing Explorer\/} ({\it RXTE\/}) \citep{whe02},
no periodic signal was detected in the hard X-ray light curves. In
contrast, strong periodic signals were detected in the soft X-ray light
curves: as shown in the upper panel of Figure 1, the power spectrum of
the soft X-ray light curve shows excess power at frequencies $\nu_0
\approx 0.0090$ Hz and $\nu_1\approx 0.11$ Hz, indicating the presence
of oscillations at periods $P_0\approx 110$ s and $P_1\approx 9.1$ s. To
investigate these oscillations more closely, we divided the soft X-ray
light curve into 47 consecutive 1 ks intervals and calculated their
power spectra. Although the individual power spectra are rather noisy,
the $\nu_1 \approx 0.11$ Hz oscillation typically appears as a single
$\Delta\nu =0.001$ Hz peak with a frequency in the range $\nu_1 =
0.109$--0.112 Hz. The mean of these 47 consecutive 1 ks power spectra
is shown in the middle panel of Figure 1. In addition to the peaks at
$\nu_0\approx 0.0090$ Hz and $\nu_1\approx 0.11$ Hz, there appears to be
a shoulder on the higher-frequency peak extending to $\nu\approx 0.12$ Hz.
To account for the observed frequency evolution, we scaled the frequency
vectors of the 47 power spectra by the instantaneous frequency of the
dominant $\nu_1$ oscillation. The mean of these scaled power spectra is
shown in the lower panel of Figure 1. Note that most of the power of
the $\nu_1$ oscillation now lies in a single frequency bin, there is no
excess power in the first harmonic $2\nu_1$ or subharmonic $\nu_1/2$
(i.e., the dominant $\nu_1$ oscillation is sinusoidal to high degree),
and the lower-frequency peak lies at $\nu_0/\nu_1\approx 0.088$.

\subsection{EUVE DS Photometer}

This is not the first time that multiple periodicities have been seen
in the short wavelength flux of SS~Cyg. In the power spectrum of the
{\it EUVE\/} DS count rate light curve of the 1994 June/July outburst of
SS~Cyg, \citet{mau97b} noted the presence of oscillations at $\nu_0\approx
0.012$ Hz, $\nu_1\approx 0.13$ Hz, and its first harmonic $2\nu_1$.
Furthermore, he noted that the harmonic peak was present in data from
only the first part of the observation, and suggested that the
low-frequency peak is due to the spin of the white dwarf. To revisit
these issues, we reexamined the DS data from the second part of the {\it
EUVE\/} observation (between 1994 June 29 $\rm 0^h49^m$ and July 3 $\rm
9^h19^m$ UT). Background-subtracted light curves were derived from 37 15
ks orbits, 37 power spectra were calculated, and the mean power spectrum
derived. As shown in the upper panel of Figure 2, peaks in this power
spectrum occur at frequencies $\nu_0\approx 0.012$ Hz and $\nu_1\approx
0.13$ Hz, indicating the presence of oscillations in the EUV flux of
SS~Cyg at periods $P_0\approx 83$ s and $P_1\approx 7.7$ s. As before,
we accounted for the observed frequency evolution from $\nu_1= 0.124$ Hz
to 0.132 Hz \citep{mau96} by scaling the frequency vectors of the 37
power spectra by the instantaneous frequency of the dominant $\nu_1$
oscillation. The mean of these scaled power spectra is shown in the lower
panel of Figure 2. Note that most of the power of the $\nu_1$ oscillation
again lies in a single frequency bin, there is no excess power in the
first harmonic $2\nu_1 $ or subharmonic $\nu_1/2$ (i.e., the dominant
$\nu_1$ oscillation is sinusoidal to high degree), and the lower-frequency
peak lies at $\nu_0/\nu_1\approx 0.096$.

\subsection{Energy Dependence of the DNOs}

Before leaving this section, we note that it is possible to use the
{\it Chandra\/} LETG data to investigate the energy dependence of the
EUV/soft X-ray oscillations of SS~Cyg. While an \AA -by-\AA \
investigation of the spectrum of the oscillations is beyond the scope of
this paper, it is a simple matter to divide the LETG soft X-ray bandpass
in two. A division at $\lambda\approx 80$ \AA \ would halve the counts
from the soft component of the X-ray spectrum of SS~Cyg, but instead we
divided the counts at $\lambda = 70$ \AA \ to approximate the bandpass of
the {\it EUVE\/} DS. As before, we constructed background-subtracted
light curves with 1 s time resolution, divided the light curves into 47
consecutive 1 ks intervals, calculated power spectra, and derived mean
power spectra. The resulting mean power spectra for the $\lambda =
42$--70 \AA \ and $\lambda = 70$--120 \AA \ bandpasses are shown
respectively in the upper and lower panels of Figure 3. This figure shows
that in the softer $\lambda = 70$--120 \AA \ bandpass the $\nu_0\approx
0.0090$ Hz and $\nu_1\approx 0.11$ Hz oscillations are slightly stronger,
while in the harder $\lambda = 42$--70 \AA \ bandpass the shoulder on the
$\nu_1\approx 0.11$ Hz oscillation appears to resolve into a distinct
peak at $\nu_2\approx 0.12$ Hz. 

\section{Discussion}

\subsection{CV DNOs and QPOs}

Summarizing the results from the previous section, we have found that
during the 1994 June/July {\it EUVE\/} observation of SS~Cyg in outburst
the EUV flux was oscillating at frequencies $\nu_0\approx 0.012$ Hz and
$\nu_1\approx 0.13$ Hz, while during the 2001 January {\it Chandra\/}
LETG observation the soft X-ray flux was oscillating at frequencies
$\nu_0\approx 0.0090$ Hz, $\nu_1\approx 0.11$ Hz, and possibly $\nu_2
\approx\nu_0+\nu_1\approx 0.12$ Hz. We note that because the frequency
of the $\nu_0$ oscillation is not constant from 1994 to 2001, it cannot
be due to the spin of the white dwarf, as suggested by \citet{mau97b}.
Instead, the near-constancy of the ratio $\nu_0/\nu_1 \approx 0.09$
suggests that the $\nu_0$ oscillation is related to the dominant $\nu_1$
oscillation.

To place these results in a broader context, we note that other
high-$\Mdot $ CVs have been observed to display multiple periodicities in
their power spectra. First, \citet{wou02} list a number of instances in
the literature when pairs of DNOs have been detected in the optical flux
of nova-like variables and dwarf novae in outburst. Second, \citet{ste01}
detected a pair of DNOs at $\nu_1 =0.0336$ Hz and $\nu_2 =0.0356$ Hz in
the optical continuum and Balmer emission line flux of V2051 Oph during
its 1998 July outburst. Although it was not possible to detect the
difference frequency $\nu_2 -\nu_1 =0.002$ Hz directly in the power
spectrum (D.\ Steeghs 2002, personal communication), the amplitude of the
oscillation varied considerably on a timescale equal to the inverse of
the difference frequency (8 min). Third, \citet{wou02} discuss a number
of instances when multiple periodicities have been detected in the optical
flux of VW~Hyi in outburst. During the decline of the 2000 February
outburst, DNOs with periods $P_{\rm DNO} =27$--37 s and QPOs with periods
$P_{\rm QPO} =400$--580 s were detected simultaneously. The ratio $P_{\rm
DNO}/ P_{\rm QPO} =0.064$--0.071, which is similar to the ratio observed
in the EUV and soft X-ray oscillations of SS~Cyg. Furthermore, in data
from the 1972 November outburst of VW~Hyi, three oscillations were
detected simultaneously: a pair of DNOs with periods $P_1=28.77$ s and
$P_2= 31.16$ s, and a QPO with a period $P_{\rm QPO}= 349$ s. The ratio
$P_2/ P_{\rm QPO}= 0.089$ and the difference frequency $1/P_1 - 1/P_2
=0.00267$ Hz, which is close, but not equal, to the QPO frequency
$1/P_{\rm QPO} =0.00287$ Hz.

\subsection{Possible Connection Between CV and LMXB QPOs}

Having established these properties of the QPOs of CVs, it is interesting
to investigate their possible connection to the QPOs of LMXBs.
\citet{kli00} provides a comprehensive review of LMXB QPOs, so it is
sufficient to note here that among neutron star binaries, the luminous Z
sources have pairs of 200--1200 Hz ``kilohertz QPOs,'' 15--60 Hz
``horizontal branch oscillations'' (HBOs), and 5--20 Hz ``normal branch
oscillations,'' while the less luminous atoll sources have 500--1250 Hz
kilohertz QPOs, as well as 20--60 Hz QPOs and broad noise components
with properties that are similar to HBOs. \citet[hereafter PBK99]{psa99}
showed that in five Z sources a tight correlation exists between the HBO
frequency $\nu_{\rm HBO}$ and the frequency $\nu _l$ of the
lower-frequency member of the pair of kHz QPOs (the ``lower kHz QPO'').
Specifically, when $\nu_l \lax 550$ Hz, $\nu_{\rm HBO}\approx 0.12\,
\nu_l^{0.95\pm 0.16}.$ Furthermore, by identifying with $\nu_{\rm HBO}$
and $\nu_l$ the frequencies of various types of peaked noise components
in atoll sources, other neutron star binaries, and black hole binaries,
PBK99 and subsequently \citet[hereafter BPK02]{bel02} extended this
correlation over nearly three orders of magnitude in frequency. As noted
by \cite{war02a, war02b}, the optical data for VW~Hyi in outburst lie
on an extrapolation of this correlation, extending it an additional two
orders of magnitude in frequency. Figure~4 shows that our EUV and soft
X-ray data for SS~Cyg in outburst also lie on an extrapolation of this
correlation, further establishing the connection between the CV and LMXB
QPOs.

This connection identifies the DNOs of CVs with the kHz QPOs of LMXBs, and
the QPOs of CVs with the HBOs (or the broad noise component identified as
such) of LMXBs. We note that the frequencies of the DNOs of CVs and the
kHz QPOs of neutron star binaries are similar in that they are comparable
to the Keplerian frequency $\nu_{\rm K}(r) = {1\over{2\pi }} (GM_\star
/r^3)^{1/2}$ at the inner edge of the accretion disk of, respectively, a
white dwarf and neutron star: $\nu_{\rm K}\le 0.14$ Hz for a $M_\star
=1\, \Msun $ white dwarf with $r\ge R_\star=5.5\times 10^8$ cm, while
$\nu_{\rm K}\lax 1570$ Hz for a $M_\star =1.4\, \Msun $ neutron star with
$r\gax 3\, R_{\rm S} =6\, GM_\star/c^2 =12.4$ km, as required by general
relativity. In addition to their frequencies, the DNOs of CVs and the kHz
QPOs of neutron star binaries are similar in that they have relatively
high coherence and high amplitudes, their frequency scales with the
inferred mass-accretion rate, and they sometimes occur in pairs.

\subsection{Implications for QPO Models}

Given the apparent connection between the QPOs of white dwarf, neutron
star, and black hole binaries, it is appropriate that we investigate the
implications for the theories of QPO formation. In the beat-frequency
models, QPOs occur at the Keplerian frequency $\nu_{\rm K}(r)$ at a
special radius in the accretion disk, and the beat of this frequency
with the stellar rotation frequency $\nu_\star $. In the magnetospheric
beat-frequency model \citep{alp85, lam85}, the HBO frequency is
identified with the beat frequency between the Keplerian frequency at the
magnetospheric radius $r_{\rm m}$ and the stellar rotation frequency:
$\nu_{\rm HBO}= \nu_{\rm K}(r_{\rm m})- \nu_\star $. In the sonic-point
beat-frequency model \citep{mil98}, some of the disk plasma makes its way
past the magnetospheric radius to the ``sonic point'' radius $r_{\rm s}$
where the disk is effectively terminated because of either radiation drag
or general relativistic corrections to Newtonian gravity. In this model,
the upper kHz QPO frequency is identified with the Keplerian frequency
at the sonic point radius: $\nu_u =\nu_{\rm K} (r_{\rm s})$, and the
lower kHz QPO frequency is identified with (one or two times) the beat
frequency between the Keplerian frequency at the sonic point radius and
the stellar rotation frequency: $\nu_l = n\, [\nu_{\rm K}(r_{\rm s})
-\nu_\star ]$, where $n=1$ or 2. Note that in this model, the stellar
rotation frequency $\nu_\star= (n\, \nu_u-\nu_l)/n$. In the relativistic
precession model \citep{ste98, ste99}, the QPO signals are due to the
fundamental frequencies of disk plasma orbiting a rapidly rotating
compact star in slightly eccentric and tilted orbits. In this model,
the upper kHz QPO frequency is identified with the Keplerian frequency:
$\nu_u=\nu_{\rm K}$, the lower kHz QPO frequency is identified with
the periastron precession frequency: $\nu_l=\nu_{\rm pp}$, and the HBO
frequency is identified with the nodal precession frequency: $\nu_{\rm
HBO}= \nu_{\rm np}$.

How does the existence of QPOs in white dwarf, neutron star, and black
hole binaries constrain the theories of QPO formation? First, as has been
pointed out by other authors, the existence of QPOs in both neutron star
and black hole binaries excludes both flavors of the beat-frequency
models (as well as {\it any\/} model requiring the presence or absence
of a stellar surface or magnetic field). Second, the DNOs of CVs exclude
the sonic-point beat-frequency model. First, there is no reason to
expect a sonic point in the inner disk of a CV: $R_\star >3\, R_{\rm S}$
and radiation drag is unimportant because the luminosity is a small
fraction of the Eddington rate ($L\approx GM_\star\Mdot/2R_\star \lax
3\times 10^{35}~{\rm erg~s^{-1}}\approx 0.002\, L_{\rm Edd}$) and because
the flow is only mildly relativistic ($v_{\rm K}\lax [GM_\star/R_\star ]
^{1/2}\lax 0.02\, c$). Second, in VW~Hyi at least, the DNO frequency
separation is not equal to one or two times the white dwarf spin
frequency. Using the {\it Hubble Space Telescope\/} ({\it HST\/}) Goddard
High Resolution Spectrograph, \citet{sio95} measured the projected
rotation velocity of the white dwarf in VW~Hyi in quiescence to be $v\,
\sin i\approx 600~\rm km~s^{-1}$. With a binary inclination $i\approx
60^\circ $ and a white dwarf mass $M_\star = 0.63~\Msun $ (hence $R_\star
= 8.4\times 10^8~\rm cm$), $\nu_\star =v/2\pi R_\star \approx 0.013$ Hz,
whereas the DNO separation frequency $\nu_u-\nu_l\approx 0.00267$ Hz.
Third, the QPOs and DNOs of CVs exclude the relativistic precession
model. First consider the nodal precession frequency, which is the sum
of the relativistic (Lense-Thirring) precession frequency $\nu_{\rm LT}$
due to frame dragging around a rapidly rotating compact star and the
classical precession frequency $\nu_{\rm cl}$ due to the quadrupole term
in the gravitational potential of an oblate star. Because $\nu_{\rm cl}$
is negative for prograde orbits, $\nu_{\rm np} \le \nu_{\rm LT}=8\pi ^2
\nu_{\rm K}^2\nu_\star I_\star / M_\star c^2$, where $I_\star\approx
0.1\, M_\star R_\star ^2$ is the moment of inertia of the star. For a
$M_\star = 1\, \Msun$ white dwarf, $\nu_{\rm K}\le 0.14$ Hz and $I_\star
\approx 6\times 10^{49}~\rm g~cm^2$, and since stability requires
$\nu_\star < \nu_{\rm K},$ $\nu_{\rm np}\lax 10^{-5}$ Hz, whereas the
observed CV QPO frequencies $\nu_{\rm QPO}\gax 10^{-3}$ Hz. Next consider
the periastron precession frequency, which is the difference between the
Keplerian and epicyclic frequencies. The latter term is $\nu_{\rm r}
\approx \nu_{\rm K}\, (1-6\, GM_\star /c^2 r)^{1/2}$ to sufficient
accuracy for a white dwarf accretor, so $\nu_{\rm pp}\approx \nu_{\rm K}
[1-(1-6\, GM_\star /c^2 r)^{1/2}]\lax 10^{-3}$ Hz, whereas the observed
CV DNO frequencies $\nu_{\rm DNO} \gax 10^{-2}$ Hz. 

In summary, assuming that the same mechanisms produce the QPOs in white
dwarf, neutron star, and black hole binaries, we find that the data
exclude the relativistic precession model and the magnetospheric and
sonic-point beat-frequency models (as well as {\it any\/} model requiring
the presence or absence of a stellar surface or magnetic field). More
promising are models, such as the transition layer model of Titarchuk
and collaborators \citep{tit98, tit99, tit02}, that interpret QPOs as
manifestations of disk accretion onto any low-magnetic field compact
object.

\subsection{Conclusion}

The results from the previous sections suggest that there is a close
relationship between the QPOs of CVs and LMXBs, with the DNOs of CVs
being the analogues of the kHz QPOs of LMXBs and the QPOs of CVs being
the analogues of the HBOs (or the broad noise component identified as
such) of LMXBs. The proposed equivalence of CV DNOs and LMXB kHz QPOs
strengthens the commonly-held belief that the frequency of the kHz QPOs
is equal to the Keplerian frequency at the inner edge of the accretion
disk because (1) the frequency ratio $\nu_{\rm DNO}/\nu_{\rm kHz}
\approx \nu_{\rm K} (R_{\rm WD})/\nu_{\rm K}(R_{\rm NS}) = (M_{\rm WD}
/M_{\rm NS})^{1/2} (R_{\rm NS}/R_{\rm WD})^{3/2}\approx 10^{-4}$ and (2)
in every case the DNO frequencies are consistent with the requirement
that $\nu_{\rm DNO}\le \nu_{\rm K}(R_{\rm WD})$ \citep{pat81, kni98}.
The proposed equivalence of CV DNOs and LMXB kHz QPOs on one hand and CV
QPOs and LMXB HBOs on the other hand can be strengthened by finding more
CVs with pairs of DNOs and more CVs in which DNOs and QPOs are observed
simultaneously. Conversely, systems known to show only QPOs could be
searched more carefully for DNOs. A interesting candidate is the dwarf
nova U~Gem in outburst, whose $\approx 0.04$~Hz QPO \citep{cor84, mas88, 
lon96} implies the existence of $\approx 0.5$~Hz (2 s) DNOs (an
oscillation period shorter than the integration times typically employed
to search for optical oscillations). For the Keplerian frequency to be
this high, the mass of the white dwarf in U~Gem must be quite high:
$M_\star\gax 1.3~\Msun $, assuming the \citet{nau72} white dwarf
mass-radius relation. This requirement is consistent with the white
dwarf mass derived by \citet{fri90} ($M_\star =1.26\pm 0.12~\Msun $),
but is inconsistent with the values derived by \citet{lon99} ($M_\star
=1.14\pm 0.07~\Msun $) and \citet{sma01} ($M_\star =1.07\pm 0.08~\Msun $).

Aside from filling in the lower-left corner of Figure 4, additional
observations of CVs are warranted because they provide unique and
quantitative tests of QPO models. Data can be obtained from the ground
in the optical and from space in the ultraviolet, EUV, and soft X-rays;
the system parameters (binary inclination, white dwarf mass, radius,
and rotation velocity) can be measured; eclipse mapping in edge-on
systems allows the sites of flux modulations to be located and dissected;
dwarf nova outbursts provide a dramatic and systematic variation in
the mass-accretion rate; diagnostic emission lines are available from
the optical through soft X-rays, and general relativistic affects are
minimal. Furthermore, technological improvements now allow {\it
spectroscopic\/} observations of the flux oscillations of CVs.
\citet{ste01} used the Keck Low Resolution Imaging Spectrograph ($\lambda
= 3600$--9200~\AA ) in continuous readout mode to study the DNOs in the
optical continuum and Balmer line flux of the eclipsing dwarf nova V2051
Oph in outburst. \citet{mar98} used the {\it HST\/} Faint Object
Spectrograph with the G160L ($\lambda = 1150$--2510~\AA ) grating to
study the wavelength dependence of the ultraviolet DNOs of the eclipsing
dwarf nova OY~Car in outburst. \citet{mau02b} used the {\it HST\/} Space
Telescope Imaging Spectrograph in time-tag mode with the E140H ($\lambda
= 1495$--1690~\AA ) echelle grating to study the ultraviolet DNOs of
SS~Cyg in outburst, while \citet{mau97a} used the {\it EUVE\/} SW
spectrometer to study the wavelength dependence of the EUV DNOs of SS~Cyg
in outburst. Such observations shed new light on the QPOs of CVs and
LMXBs.

\section{Summary}

We have shown that during the 1994 June/July {\it EUVE\/} observation of
SS~Cyg in outburst the EUV flux was oscillating at frequencies $\nu_0
\approx 0.012$ Hz and $\nu_1\approx 0.13$ Hz, while during the 2001
January {\it Chandra\/} LETG observation the soft X-ray flux was
oscillating at frequencies $\nu_0\approx 0.0090$ Hz, $\nu_1\approx 0.11$
Hz, and possibly $\nu_2 \approx\nu_0+\nu_1\approx 0.12$ Hz. These
data, combined with the optical data of \citet{wou02} for VW~Hyi, extend
the PBK99 and BPK02 $\nu_{\rm high}$--$\nu_{\rm low}$ correlation for
neutron star and black hole LMXBs nearly two orders of magnitude in
frequency, with $\nu_{\rm low} \approx 0.08\, \nu_{\rm high}$. This
correlation identifies the DNOs of CVs with the kHz QPOs of LMXBs, and
the QPOs of CVs with the HBOs (or the broad noise component identified
as such) of LMXBs. Assuming that the same mechanisms produce the QPOs
in white dwarf, neutron star, and black hole binaries, we find that the
data exclude the relativistic precession model and the magnetospheric
and sonic-point beat-frequency models (as well as {\it any\/} model
requiring the presence or absence of a stellar surface or magnetic
field). Additional observations of CVs can help establish the proposed
equivalence of CV and LMXB QPOs, and will provide unique and quantitative
tests of QPO models.

\acknowledgments

Our satellite observations of SS~Cyg were made possible by the optical
monitoring and alerts provided by the members, staff (particularly E.\
Waagen), and director, J.\ Mattei, of the American Association of
Variable Star Observers. The {\it EUVE\/} observation was made possible
by the efforts of {\it EUVE\/} Project Scientist R.\ Malina, Science
Planner M.\ Eckert, the staff of the {\it EUVE\/} Science Operations
Center at the Center for EUV Astrophysics, and the Flight Operations
Team at Goddard Space Flight Center. The {\it Chandra\/} observation
was made possible by the efforts of {\it Chandra\/} X-Ray Observatory
Center Director H.\ Tananbaum, Mission Planner K.\ Delain, and the {\it
Chandra\/} Flight Operations Team at MIT. We acknowledge T.\ Belloni for
kindly supplying us with the neutron star and black hole binary data 
shown in Figure 4; L.\ Titarchuk, K.\ Wood, and D.\ Steeghs for helpful
discussions; and the anonymous referee for comments which improved the
clarity of the manuscript. Support for this work was provided by NASA
through {\it Chandra\/} Award Number GO1-2023A issued by the {\it
Chandra\/} X-Ray Observatory Center, which is operated by the Smithsonian
Astrophysical Observatory for and on behalf of NASA under contract
NAS8-39073. This work was performed under the auspices of the U.S.\
Department of Energy by University of California Lawrence Livermore
National Laboratory under contract No.\ W-7405-Eng-48.

\clearpage 


\clearpage 


\begin{figure}
\figurenum{1}
\epsscale{0.5}
\plotone{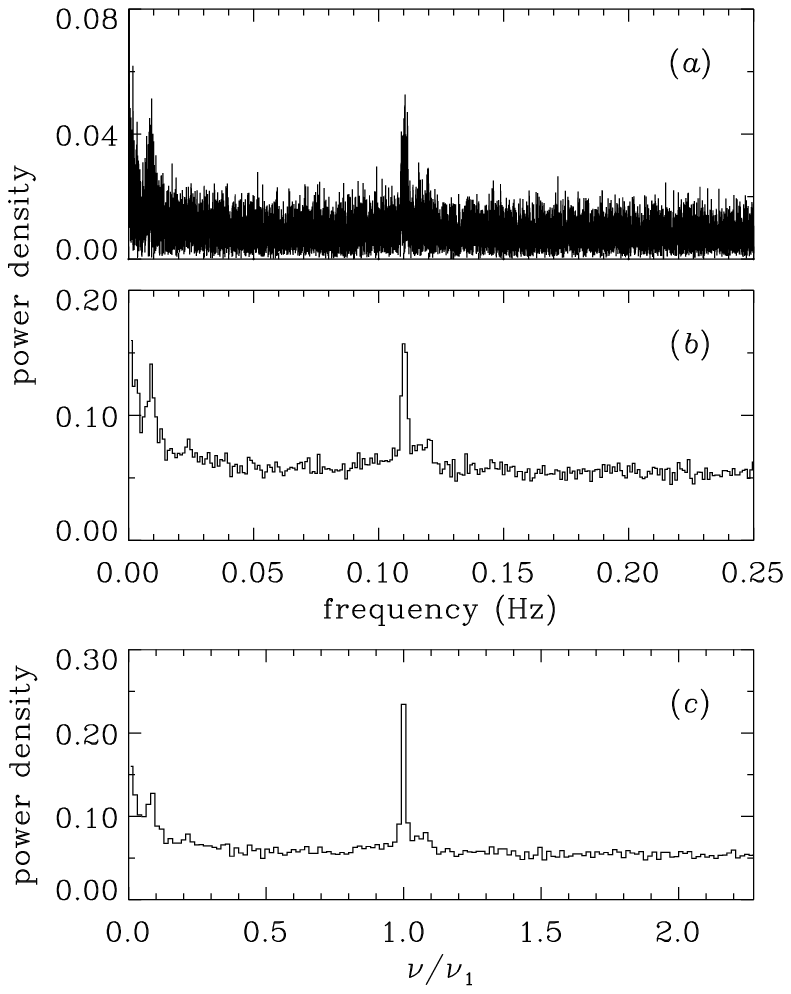}
\caption{Power spectra of {\it Chandra\/} LETG $\lambda =42$--120 \AA \
count rate light curves of SS~Cyg in outburst. ({\it a\/}) Power spectrum
of 47 ks of data binned to 1 s time resolution. ({\it b\/}) Mean power
spectrum of 47 consecutive 1 ks light curves. ({\it c\/}) Mean power
spectrum of the 47 1 ks light curves after scaling by the varying
frequency of the $\nu_1\approx 0.11$ Hz oscillation. Note the simultaneous
presence of oscillations at $\nu_0\approx 0.0090$ HZ, $\nu_1\approx 0.11$
Hz ($\nu_0/\nu_1\approx 0.088$), and possibly a third at $\nu_2\approx
\nu_0+\nu_1\approx 0.12$ Hz.}
\end{figure}

\begin{figure}
\figurenum{2}
\epsscale{0.5}
\plotone{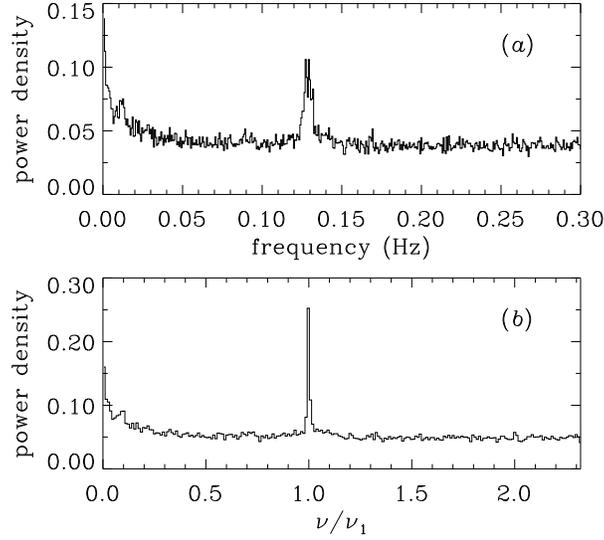}
\caption{Power spectra of {\it EUVE\/} DS count rate light curves of
SS~Cyg in outburst. ({\it a\/}) Mean power spectrum of 37 1.5 ks light
curves. ({\it b\/}) Mean of the 37 1.5 ks light curves after scaling by
the varying frequency of the $\nu_1\approx 0.13$ Hz oscillation. Note
the simultaneous presence of oscillations at $\nu_0\approx 0.012$ Hz and
$\nu_1\approx 0.13$ Hz ($\nu_0/\nu_1\approx 0.096$).}
\end{figure}

\begin{figure}
\figurenum{3}
\epsscale{0.5}
\plotone{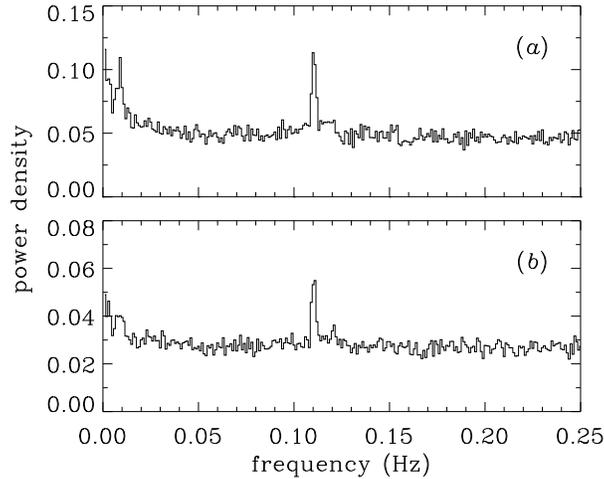}
\caption{Mean power spectra of 47 consecutive 1 ks {\it Chandra\/} LETG
count rate light curves of SS~Cyg in outburst in the ({\it a\/}) soft
$\lambda = 70$--120 \AA \ and ({\it b\/}) hard $\lambda = 42$--70 \AA \ 
bandpasses. Note that the $\nu_0\approx 0.0090$ Hz and $\nu_1\approx
0.11$ Hz oscillations are slightly stronger in the soft bandpass, and that
there is evidence for a distinct oscillation at $\nu_2\approx\nu_0+\nu_1
\approx 0.12$ Hz in the hard bandpass.}
\end{figure}

\begin{figure}
\figurenum{4}
\epsscale{0.5} 
\plotone{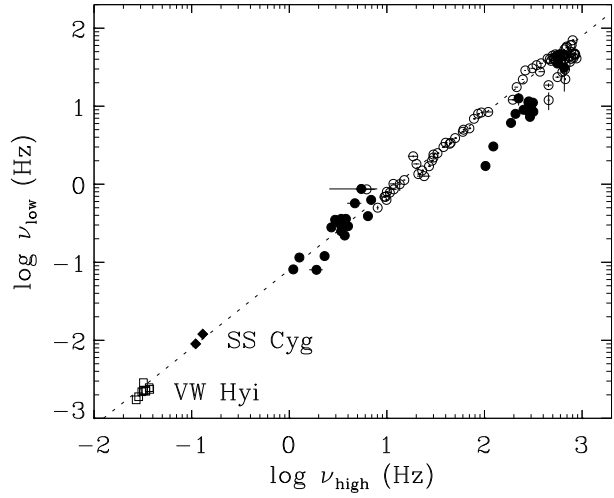}
\caption{$\nu_{\rm high}$--$\nu_{\rm low}$ correlation for neutron
star binaries ({\it open circles}), black hole binaries ({\it filled
circles}), and the white dwarf binaries SS~Cyg ({\it filled diamonds\/})
and VW~Hyi ({\it open squares\/}). Neutron star and black hole binary
data are from Figure 12 of BPK02, and were kindly supplied by T.\
Belloni. Dotted line drawn through the points is $\nu_{\rm low} =0.08
\, \nu_{\rm high}$.}
\end{figure}

\end{document}